\documentclass[twocolumn]{aastex62}
\usepackage{float}
\usepackage{afterpage}

\shorttitle{Draft}
\shortauthors{Sarkar et al.}

\begin{document}

\title{\textbf{An Observationally Constrained Analytical Model for Predicting the Magnetic Field Vectors of ICMEs at 1 AU}}

\correspondingauthor{Ranadeep Sarkar}
\email{ranadeep@prl.res.in}
 \author[0000-0002-0786-7307]{Ranadeep Sarkar}
 \affil{Udaipur Solar Observatory, Physical Research Laboratory,
 Badi Road, Udaipur 313001, India} 
\affil{Discipline of Physics, Indian Institute of Technology, Gandhinagar 382355, India} 
 \\

 \author{Nat Gopalswamy}
 \affiliation{NASA Goddard Space Flight Center, Greenbelt, MD 20771, USA}

 \author{Nandita Srivastava}
\affiliation{Udaipur Solar Observatory, Physical Research Laboratory,
Badi Road, Udaipur 313001, India}

\begin{abstract}

We report on an observationally constrained analytical model, the INterplanetary Flux ROpe Simulator (INFROS), for predicting the magnetic-field vectors of coronal mass ejections (CMEs) in the interplanetary medium. The main architecture of INFROS involves using the near-Sun flux rope properties obtained from the observational parameters that are evolved through the model in order to estimate the magnetic field vectors of interplanetary CMEs (ICMEs) at any heliocentric distance. We have formulated a new approach in INFROS to incorporate the expanding nature and the time-varying axial magnetic field-strength of the flux rope during its passage over the spacecraft. As a proof of concept, we present the case study of an Earth-impacting CME which occurred on 2013 April 11. Using the near-Sun properties of the CME flux rope, we have estimated the magnetic vectors of the ICME as intersected by the spacecraft at 1 AU. The predicted magnetic field profiles of the ICME show good agreement with those observed by the in-situ spacecraft. Importantly, the maximum strength ($10.5 \pm 2.5$ nT) of the southward component of the magnetic field (Bz) obtained from the model prediction, is in agreement with the observed value (11 nT). Although our model does not include the prediction of the ICME plasma parameters, as a first order approximation it shows promising results in forecasting of Bz in near real time which is critical for predicting the severity of the associated geomagnetic storms. This could prove to be a simple space-weather forecasting tool compared to the time-consuming and computationally expensive MHD models.

\end{abstract}

\keywords{coronal mass ejections (CMEs), flares, solar-terrestrial relations\\
}

%

%

\section{Introduction}
Coronal mass ejections (CMEs) are powerful expulsions of gigantic clouds of magnetized plasma that routinely erupt from the Sun and propagate out through the solar system. When such an eruption is directed toward the Earth with high speed and its north-south magnetic field component (Bz) is directed towards the south, an intense magnetic storm occurs upon the impact of the CME on Earth's magnetosphere \citep{Wilson1987,Tsurutani,Gonzalez,Huttunen,Yurchyshyn,Gopalswamy2008}. The storm can occur when the interplanetary flux rope (FR) and/or the sheath between the FR and the associated shock has southward Bz. Therefore, a prior knowledge of the strength and orientation of the magnetic field embedded in the FR is required in order to forecast the severity of geomagnetic storms caused by CMEs.

Several modeling efforts have been made in order to predict Bz at 1 AU \citep{Odstrcil,Shen,Savani2015,Jin,Kay2017,Mostl}. However, due to the complexity of the Sun-Earth system in a time-dependent heliospheric context, the semi-analytical and global MHD models are usually unable to reproduce the strength and orientation of the magnetic field vectors observed by the in-situ spacecraft. 
The FR from Eruption Data (FRED) technique published recently can be used to obtain the magnetic properties of the near-Sun coronal FRs from the photospheric magnetic flux under post eruption arcades and the geometric properties of the FR obtained from the fitting of white-light coronagraphic structures \citep{Gopalswamy_fred, Gopalswamy2018}. In this work we have developed an analytical model,  the INterplanetary Flux ROpe Simulator (INFROS), that utilizes FRED parameters as realistic inputs and evolves those parameters in real time to predict the magnetic field vectors of interplanetary coronal mass ejections (ICMEs) reaching at Earth.

Apart from using the realistic inputs, we have formulated a new approach in our model to incorporate the expanding nature and the time-varying axial magnetic field-strength of the FR during its passage over the spacecraft. In contrast to existing models \citep{Savani2015, Kay2017,Mostl} our approach is unique in that it does not involve any free parameters like the dimension, axial field strength, time of passage and the speed of ICME at 1 AU. Therefore, INFROS is the first such model which uses the realistic inputs to predict the magnetic field vectors of ICMEs without involving any free parameters.

In principle, INFROS can be used to estimate the magnetic field vectors of ICMEs at any heliocentric distance. Importantly, the prediction of magnetic field vectors of Earth-reaching ICMEs at 1 AU is crucial for space-weather forecasting. Therefore, in this paper we have considered this heliocentric distance as 1 AU for explaining the development of the model.

We have organized this article as follows. The observational reconstruction techniques of the near-Sun FR parameters are discussed in Section \ref{first}. In Section \ref{model}, we have described the model architecture developed to predict the ICME vector profiles at 1 AU. We validate our model for a test case in Section \ref{model_validation}. Finally,  we summarize our results and discuss their implications for space-weather forecasting in Section \ref{discussion}.

\section{Near-Sun observations of flux rope properties}\label{first}
We determine the geometric and magnetic properties of the near-Sun FRs using the FRED technique as described in this section.  
\subsection{Geometrical properties}\label{sec:gcs}
We determine the three-dimensional morphology and the propagation direction of CMEs by using the graduated cylindrical shell (GCS) \citep{GCS} model. This model fits the geometrical structure of CMEs as observed by white-light coronagraphs such as the Large Angle and Spectrometric Coronagraph (LASCO) \citep{LASCO} on board the Solar and Heliospheric Observatory (SOHO) \citep{SOHO} mission, and Sun Earth Connection Corona and Heliospheric Investigation (SECCHI) \citep{SECCHI} on board the Solar Terrestrial Relations Observatory (STEREO) \citep{STEREO} mission. Using the GCS model, we obtain the propagation longitude ($\phi$) and latitude ($\theta$), half-angular width ($\beta$), aspect ratio ($\kappa$), tilt angle ($\gamma$) with respect to the solar equator and the leading-edge height (h) of the CME FR. 

The parameter $\kappa$ constrains the rate of expansion of the CME FR under the assumption of self-similar expansion. Therefore, the cross-sectional radius (r) of the self-similarly expanding FR at any heliocentric distance R $(=h-r)$, can be obtained using the relation, r$=\kappa$h/(1+$\kappa$). On the other hand, the length (L) of the flux-rope can be estimated from the relation, $L=2 \beta R$, where $2 \beta$ is the separation angle between the two legs of the CME in radian.      
\subsection{Magnetic properties} 
Observational approaches to determine the three magnetic parameters which completely define any force-free FR are discussed as follows.

\subsubsection{Axial field strength (B$_0$)}\label{fred}
Several studies have shown that the azimuthal (poloidal) flux of magnetic FRs formed due to the reconnection is approximately equal to the low-coronal reconnection flux, which can be obtained either from the photospheric magnetic flux underlying the area swept out by the flare ribbons \citep{Longcope, Qiu} or the magnetic flux underlying the post eruption arcades \citep{Gopalswamy1}. Combining the geometrical parameters of the FR obtained from the GCS fitting as discussed in Section \ref{sec:gcs} with the estimation of reconnected magnetic flux, \citet{Gopalswamy2018} introduced the FRED model which shows that the axial magnetic-field strength of the FR can be determined using a constant alpha force-free FR model \citep{Lundquist}. Thereby, we obtain the magnetic field strength (B$_0$) along the FR axis using the relation \citep{Gopalswamy_fred, Gopalswamy2018},
\begin{equation}\label{axial_field_strength}
B_0=\frac{\phi_px_{01}}{Lr} 
\end{equation}
where $\phi_p$ is the azimuthal magnetic flux taken as the reconnection flux, $x_{01}$ ($=$ 2.4048) is the first zero of the Bessel function $J_0$, L is the length and r is the cross-sectional radius of the FR.
\subsubsection{Direction of the axial magnetic field and the sign of helicity}
In order to determine the direction of the axial magnetic field and the helicity sign (chirality) associated with the FR, we first apply the hemispheric helicity rule to the source active region of the CME as first order approximation \citep{Pevtsov,Bothmer}. However, the statistical studies by \citet{LiuHoeksema} show that the hemispheric rule is followed only in 60\% of cases. Therefore, in order to confirm the chirality and the axial orientation of the FRs we use other signatures such as pre-flare sigmoidal structures \citep{Rust_kumar}, J-shaped flare ribbons \citep{Janvier_2014}, coronal dimmings \citep{Webb,Thompson1998,Gopalswamy2018c}, coronal cells \citep{Sheeley} or filament orientations \citep{Hanaoka}. Analyzing the locations of the two core dimming regions or the two ends of the pre-flare sigmoidal structure, one can identify the locations of the two foot points  of the FR. Thereafter, the locations of the FR foot points can be overlaid on the line-of-sight magnetogram to determine in which magnetic polarities the FR is rooted \citep{Palmerio2017}. Once the direction of the axial field is determined, one can confirm the helicity sign (chirality) from the positive and negative polarities that are divided by the neutral line \citep{Bothmer,Marubashi2015,Gopalswamy_fred}.

CMEs may undergo rotation in the lower corona depending on the amount of sigmoidality or the skew present in the associated pre-eruptive FR structure \citep{Lynch}. Therefore, one can get mismatch between the FR orientation determined from the on-disk observations and the tilt angle of the CME obtained from the GCS fitting. Moreover, considering an uncertainty of $\pm$ 20$^{\circ}$ in determining the on-disk axis orientation \citep{Palmerio2018} and $\pm$ 10$^{\circ}$ in determining the GCS tilt angle \citep{Thernisien2011}, one may obtain difference in angle upto $\pm$ 30$^{\circ}$ between the GCS tilt and the on-disk axis orientation, in absence of any significant rotation of the associated CME.   Therefore, in order to resolve the 180$^{\circ}$ ambiguity in determining the FR axis orientation from the GCS tilt, we consider the smallest angle ($<180 ^{\circ}$) between the on-disk and the GCS axis orientation. In this way we can determine the direction of axial magnetic field of the CME observed in coronagraphic field-of-view.

\begin{figure*}[!t]
\centering
\includegraphics[width=.49\textwidth]{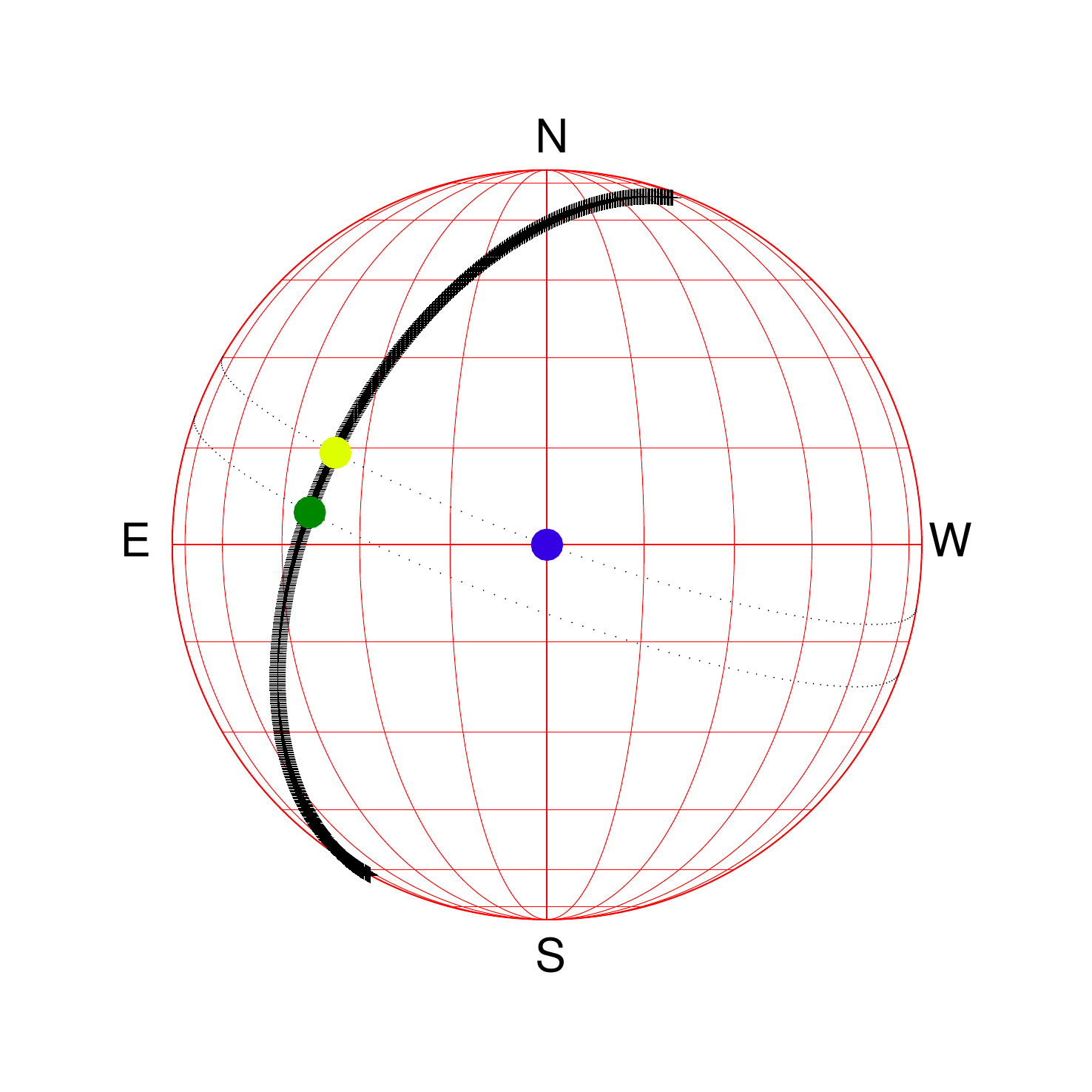}
\includegraphics[width=.49\textwidth]{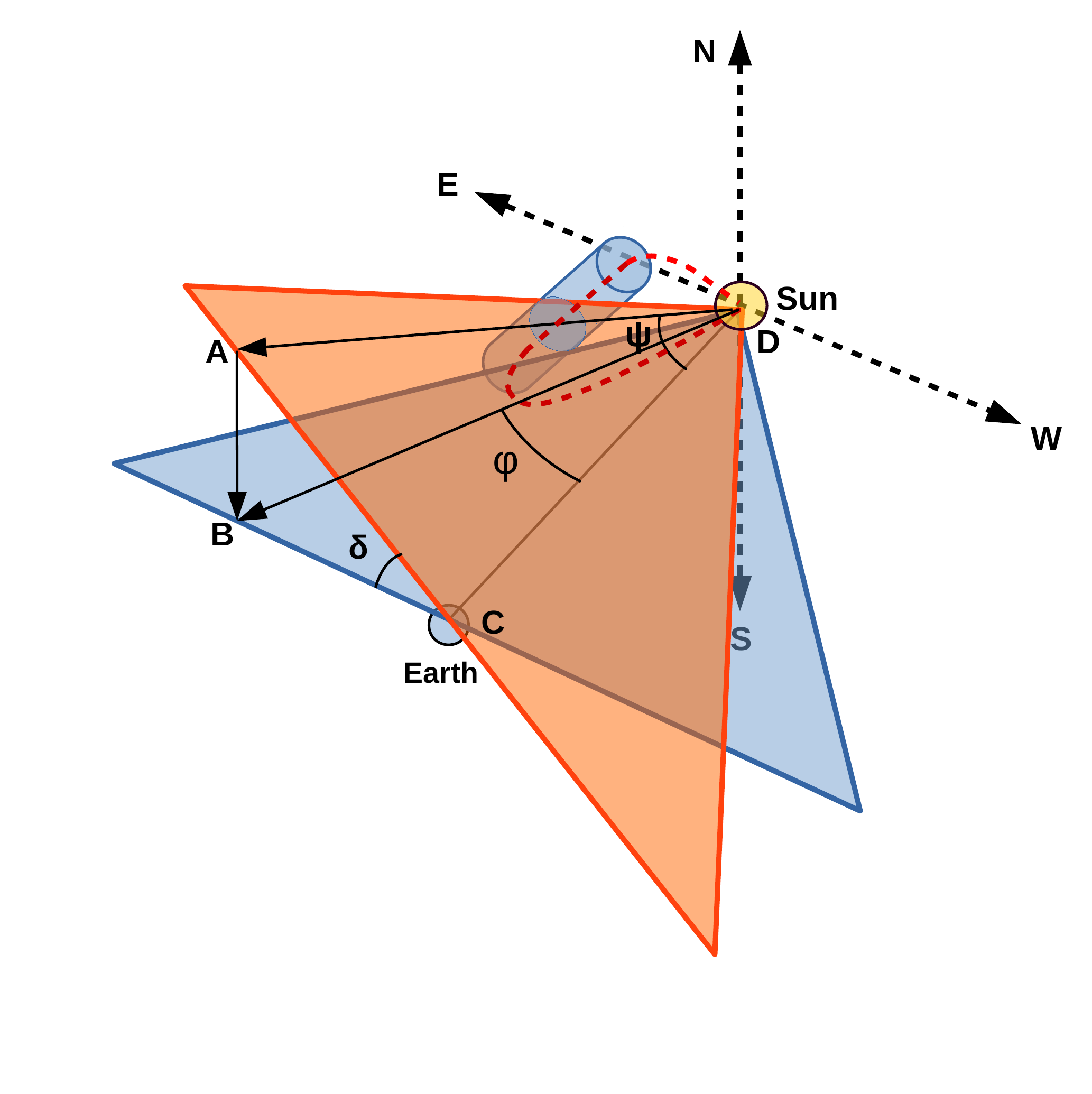}
\caption{\textit{Left panel}: The black solid line denotes the projected CME axis on the solar disc. Solar grids are shown in red with 15$^{\circ}$ intervals in both longitude and latitude. The projected location of Earth is indicated by the blue dot; the green dot marks the center of the CME axis. The yellow dot marks the location on the CME axis which is intersected by the black dotted line connecting the blue dot and perpendicular to the CME axis. \textit{Right panel}: Schematic picture of an MC propagating through the interplanetary space in between the Sun and Earth. The red dashed line indicates the axis of the MC. Locations of the Sun and Earth are indicated by the points D and C respectively. The blue plane depicts the ecliptic plane, whereas the orange one is perpendicular to the MC axis and passes through the Sun-Earth line (CD). The MC axis is tilted by an angle $\gamma$ with respect to the ecliptic plane. Therefore, the plane (orange) perpendicular to the MC axis makes an angle $\delta (=90^{o}-\gamma)$ with respect to the ecliptic plane (blue). The line connecting A and D lies on the orange plane and intersects the MC axis along the longitudinal direction $\phi$ (longitude of the yellow dot marked in the right panel) with respect to the Sun-Earth line. BD is the projection of line AD on the ecliptic plane (blue). The angle ($\psi$) between AD and CD, denotes the separation angle  between the MC axis and the Sun-Earth line.}
\label{impact_distance}
\end{figure*}

\section{Modeling the interplanetary flux ropes using the near-Sun observations}\label{model}
We track the evolution of the near-Sun FR properties using the analytical model (INFROS) and estimate the magnetic field vectors of the associated interplanetary FRs known as the magnetic clouds (MCs). Notably, the MCs are a subset of ICMEs which show enhanced magnetic fields with a smooth rotation in the direction of field vectors, and low proton temperature during its passage over the in-situ spacecraft \citep{Burlaga}. On the other hand, the ICMEs which lack the MC signatures in their in-situ profile are known as non-cloud ejecta. The internal magnetic field structure of those ICMEs does not resemble that of a magnetic FR. However, it is important to note that all ICMEs may have the FR structures, but their in-situ observations may lack that coherent magnetic structure depending on the path of the observing spacecraft \citep{Kim2013, Gopalswamy2006ssrv}. Therefore, similar to the existing semi-analytical and analytical models \citep{Savani2015, Kay2017,Mostl}, INFROS is applicable for all ICMEs in general, but can be validated only for those ICME events which show MC signatures in their in-situ profile.

As significant deflection and rotation of CMEs generally occur very close (less than 10 $R_S$) to the Sun \citep{Kay,Lynch}, we assume that the propagation direction and the axis-orientation of the CME obtained from the GCS fitting at approximately 10 $R_S$ are maintained throughout its evolution from the Sun to Earth. We also do not consider any CME-CME interaction in the interplanetary space which may change the propagation trajectory of the CME. Assuming that the CMEs expand in a self-similar \citep{Subramanian_2014, good2019self‐similarity, Vr_nak_2019} way during its interplanetary propagation, we estimate the geometrical parameters of the CME upon its arrival at 1 AU. Using the conservation principle of the magnetic flux and helicity, we determine the magnetic properties of the FR when it is intersected by the spacecraft at 1 AU. Finally, incorporating those estimated geometrical and magnetic parameters of the FR in a constant alpha force-free FR solution \citep{Lundquist} we estimate the expected magnetic vector profiles of Earth-impacting ICMEs. The detailed description of the INFROS model is as follows.

\subsection{Estimating the impact distance}
In order to estimate which part of the ICME will be intersected by the observing spacecraft at 1 AU, it is important to first determine the impact distance ($d$) that is the closest distance between the MC axis and the location of the spacecraft. According to the geometry illustrated in Figure \ref{impact_distance}, we can write  
\begin{equation}
BC=R_{SE}\times{tan\phi}
\end{equation}

where, $R_{SE}$ is the distance between Sun and Earth and $\phi$ is the longitudinal direction of the line DA. As the plane perpendicular to the MC axis is tilted by an angle $\delta$, we can further write
\begin{equation}\label{AC}
AC=\frac{BC}{cos\delta}=R_{SE}\times\frac{tan\phi}{cos\delta}
\end{equation}
Using the value of AC from Equation \ref{AC}, we can obtain the minimum separation angle $\psi$ between the axis of the MC and the Sun-Earth line from the following relation
\begin{equation}
tan\psi=\frac{AC}{R_{SE}}=\frac{tan\phi}{cos\delta}
\end{equation}
After determining the value of $\psi$, the impact distance (d) of the MC at any helio-centric distance (R) along the Sun-Earth line can be obtained from the following equation
\begin{equation}
d=R\times{sin\psi}=R\times{sin(tan^{-1}\frac{tan\phi}{cos\delta})}
\label{p_impact}
\end{equation}

\begin{figure}[!h]
\centering
\includegraphics[width=.45\textwidth]{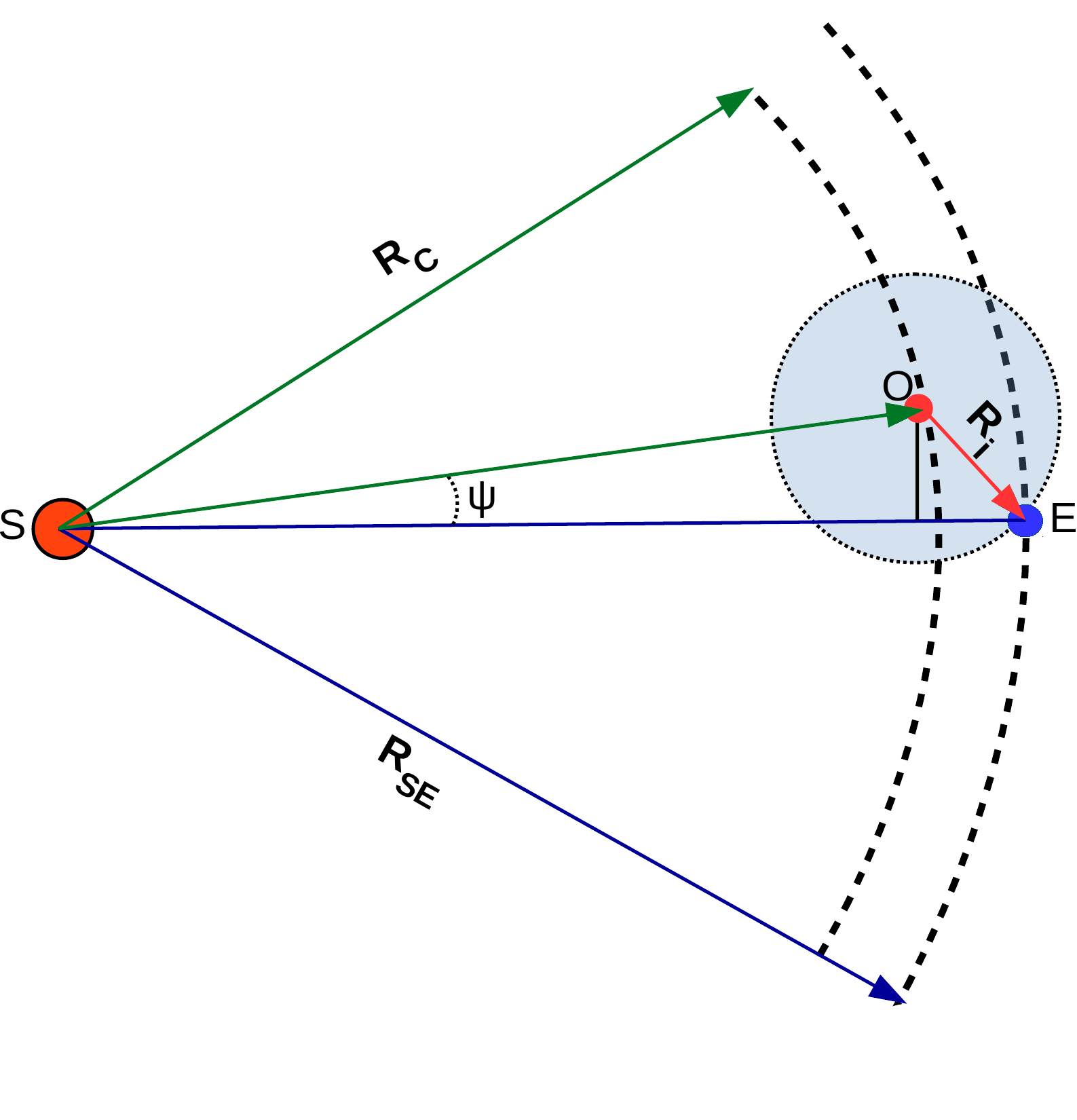}
\caption{Schematic picture of the MC cross-section on the plane (the orange plane as depicted in Figure \ref{impact_distance}) perpendicular to the MC axis. The MC axis is pointing out of the plane at point O. The angle $\psi$ denotes the separation angle between the MC axis and the Sun-Earth line. R$_C$ is the radial distance of the MC axis from the Sun-center and R$_i$ is the radius of cross-section when the spacecraft just encounters the arrival of MC}
\label{initial_radius}
\end{figure}

\subsection{Cross-sectional radius of the flux rope when the spacecraft just encounters the arrival of MC}

In order to infer the axial field-strength of the MC from the conservation of magnetic flux, we need to estimate its cross-sectional area during its passage over the spacecraft. Figure \ref{initial_radius} depicts a schematic picture of an MC cross-section when the spacecraft just encounters its arrival. According to the geometry as illustrated in Figure \ref{initial_radius}, we can write

\begin{equation}
R_c\times{cos{\psi}}+\sqrt{R_i^2-{{R_c}^2\times{sin^2{\psi}}}}=R_{SE}
\label{initial_radius_eq}
\end{equation}
where, R$_C$ is the radial distance of the MC axis from the Sun-center, $\psi$ is the separation angle between the MC axis and the Sun-Earth line and R$_i$ is the radius of cross-section of the MC. Assuming that the CME has evolved self-similarly between Sun and Earth, we can replace R$_c$ in Equation \ref{initial_radius_eq} using the relation $R_i={\kappa}R_c$, where $\kappa$ is the aspect ratio of the CME FR obtained from the observations as discussed in \ref{sec:gcs}. Thereby, we can estimate the initial radius of the FR cross-section upon its arrival at Earth using the following equation

\begin{equation}
R_i=\frac{\kappa\times{R_{SE}}}{cos{\psi}+\sqrt{\kappa^2-sin^2{\psi}}}
\label{ini_rad_eq_2}
\end{equation}

For, $\psi$=0, Equation \ref{ini_rad_eq_2} reduces to Equation \ref{ini_rad_eq_3}, which is the scenario when the spacecraft passes through the center of the FR cross-section. 

\begin{equation}
R_i=\frac{\kappa\times{R_{SE}}}{1+\kappa}
\label{ini_rad_eq_3}
\end{equation}
\subsection{Self-similar approach to incorporate the flux rope expansion during its passage through the spacecraft}\label{SSE}
\begin{figure}[!b]
\centering
\includegraphics[width=.48\textwidth]{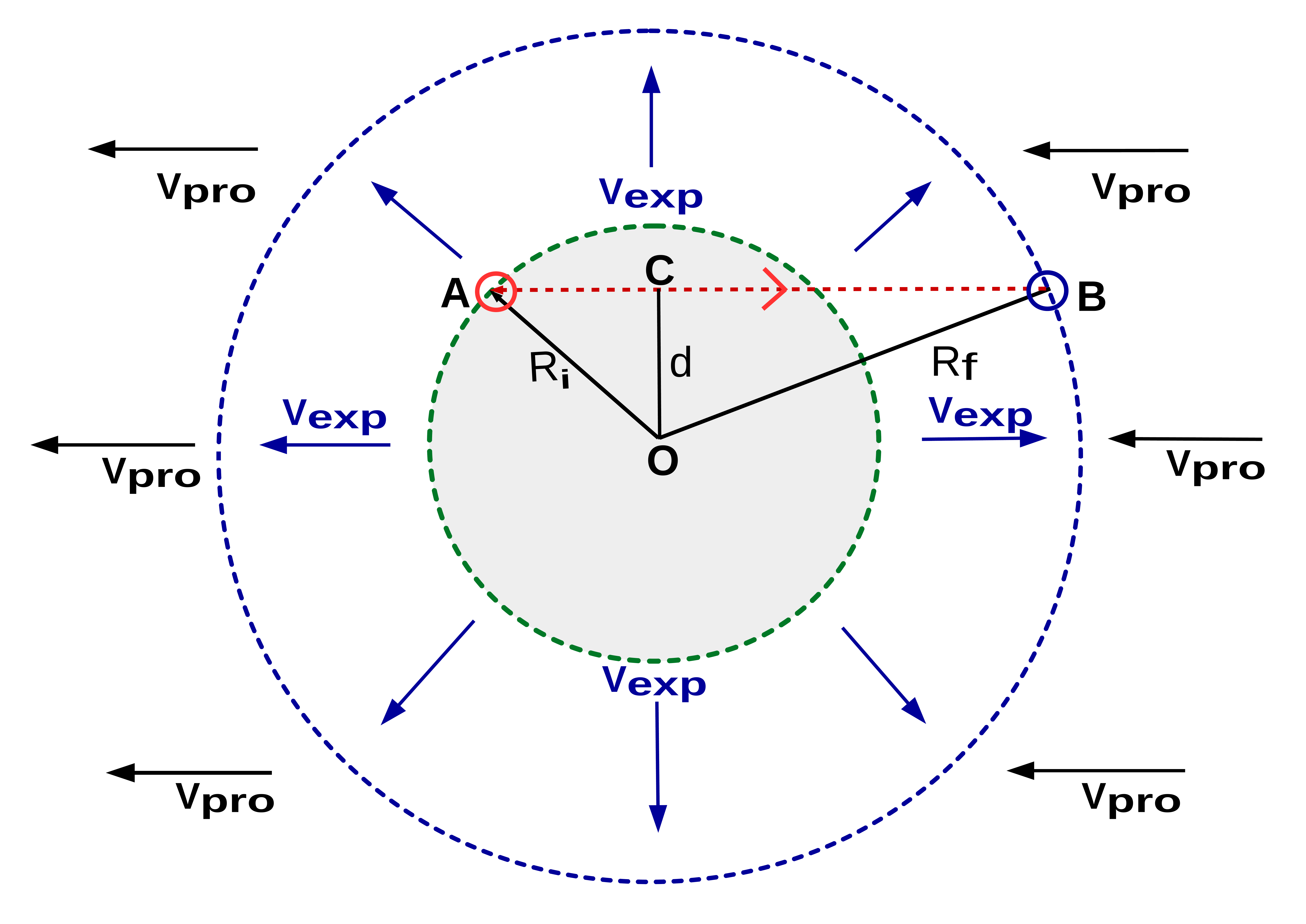}
\caption{Schematic picture of the cross-section of an expanding FR as it passes over the spacecraft with a propagation speed $V_{pro}$ and expansion speed $V_{exp}$. The black arrows denote the direction of the MC propagation, whereas the blue arrows represent the isotropic expansion of the MC. The spacecraft intersects the MC at an impact distance ``d" denoted by OC. The gray shaded region encircled by the green dashed line denotes the initial boundary of the FR with cross-sectional radius $R_i$ when the spacecraft just encounters the arrival of MC at point A marked by the red circle. The red dotted line illustrates the trajectory of the spacecraft from A to B inside the expanding MC. $R_f$ is the final radius of the MC cross-section encircled by the blue dashed line when the spacecraft encounters the end-boundary of the MC.}
\label{expansion}
\end{figure}

Figure \ref{expansion} depicts the spacecraft trajectory inside the MC assumed to expand isotropically with expansion speed $V_{exp}$. The MC axis propagates with a speed $V_{pro}$ along the direction depicted by the black arrows in Figure \ref{expansion}. Therefore, in the FR frame of reference, the spacecraft traverses from the point A (lies on the front-boundary of the MC) to the point B (lies on the rear boundary of the MC) with a speed $V_{pro}$. If $t_p$ is the travel time for the spacecraft to complete the path AB, we can write 
  \begin{equation}
  \sqrt{R_{i}^2-d^2}+\sqrt{R_{f}^2-d^2}=v_{pro}\times{t_{p}}
  \label{vpro}  
  \end{equation}

where, $R_i$ and $R_f$ are the cross-sectional radius of the front and rear boundary of the MC respectively and `d' is the impact distance of the spacecraft from the MC axis. By the time ($t_p$) the spacecraft traversed the path AB, the cross-sectional radius of the MC increased from $R_i$ to $R_f$ with the expansion speed $V_{pro}$. Therefore, we can write
  
  \begin{equation}
  R_{f}-R_{i}=v_{exp}\times{t_{p}}
  \label{vexp}
  \end{equation}

Considering a general case, where the MC axis takes $t_{travel}$ time to traverse a distance $R_{tip}$ with a speed $v_{pro}$, we can write 
  
  \begin{equation}
  R_{tip}=v_{pro}\times{t_{travel}}
  \label{rtip}
  \end{equation}
During the time $t_{travel}$, as the cross-sectional area of the MC also expands with a speed $v_{exp}$, the final radius of the MC cross-section after $t_{travel}$ can be written as   
  \begin{equation}
  R_{cross}=v_{exp}\times{t_{travel}}
  \label{rcross}
  \end{equation}
Using the properties of self-similar expansion, $R_{cross}$ and $R_{tip}$ can be related as  $R_{cross}={\kappa}R_{tip}$. Therefore, using the Equations \ref{rtip} and \ref{rcross}, we can relate $v_{pro}$ and $v_{exp}$ through the following relation
  
  \begin{equation}    
  \frac{R_{cross}}{R_{tip}}=\frac{v_{exp}}{v_{pro}}=\kappa
  \label{ratio}
  \end{equation}
Using the Equations \ref{vpro}, \ref{vexp} and \ref{ratio}, we can write  
  
  \begin{equation}
  \frac{R_{f}-R_{i}}{\sqrt{R_{i}^2-d^2}+\sqrt{R_{f}^2-d^2}}=\frac{v_{exp}}{v_{pro}}=\kappa
  \label{rf1}
  \end{equation}
In Equation \ref{rf1}, $R_i$, d and $\kappa$ are the known parameters.    $R_i$ is obtained from Equation \ref{ini_rad_eq_2}, impact distance `d' is obtained from the Equation \ref{p_impact} and the value of $\kappa$ is obtained from the observations as discussed in the Section \ref{sec:gcs}. Rewriting the Equation \ref{rf1}, we get the following quadratic equation of $R_f$   
  
  \begin{equation}
  R_{f}^2+b\times{R_f}+c=0.
  \label{rf}
  \end{equation}
  where, $$ b= \frac{2\times({R_i+\kappa\times{\sqrt{R_{i}^2-d^2}}})}{\kappa^2-1}$$
         $$ c= \frac{({R_i+\kappa\times{\sqrt{R_{i}^2-d^2}}} \; )^{2}-d^2\times{\kappa^2}}{1-\kappa^2} $$

Therefore, solving the Equation \ref{rf} we can estimate the final radius ($R_f$) of the expanding FR when the spacecraft encounters the rear-boundary of the MC. After estimating $R_i$ (initial radius of the MC front-boundary), $R_f$ (final radius of the MC rear-boundary) and `d' (impact distance), we can estimate the path AB as depicted in Figure \ref{expansion}.
In order to capture the full expansion profile of the MC, next we need to determine the cross-sectional radius of the expanding FR at any distance $x$ traversed by the spacecraft throughout the path AB (Figure \ref{expansion}). Let us consider, at any time t ($0{\le} t {\le}t_p$) the SC traverses a distance x with a speed $v_{pro}$ along AB in the frame of reference attached to the MC axis. Therefore we can write, 

  \begin{equation}
  x=v_{pro}\times{t}
  \label{x1}
  \end{equation}
During the time t, the cross-sectional radius of the FR increases from $R_i$ to $R_t$ with a speed $v_{exp}$. Therefore we can write   
  \begin{equation}
  R_t-R_i=v_{exp}\times{t}
  \label{x2}
  \end{equation}
Using the Equations \ref{rf1}, \ref{x1} and \ref{x2}, we can further write  
  \begin{equation}
  \frac{R_t-R_i}{x}=\frac{v_{exp}}{v_{pro}}=\kappa
  \label{x3}  
  \end{equation}
Rewriting the Equation \ref{x3} we get   
  \begin{equation}
  R_t=R_i+\kappa\times{x}
  \label{x4}
  \end{equation}
Therefore at any distance $x$ along the path AB (Figure \ref{expansion}), we can estimate the cross-sectional radius ($R_i{\le} R_t {\le}R_f$) of the expanding FR using the Equation \ref{x4}. It is noteworthy that we have started our formulation with the unknown parameters $V_{exp}$, $V_{pro}$ and $t_{p}$ (see Equations \ref{vpro} and \ref{vexp}) and finally arrived to the Equations  \ref{rf} and \ref{x3}, which are independent of the aforementioned variables. This is the major advantage of this formulation as we have incorporated the FR expansion in such a way so as to get rid of the free or unknown parameters like the expansion speed ($V_{exp}$), propagation speed ($V_{pro}$) and the time of passage ($t_{p}$) of the ICMEs at 1 AU.

\subsection{Estimating the final magnetic field profiles of the MC at 1 AU using a cylindrical flux rope solution}
It is expected that the FR axial field strength ($B_0$) will decrease as the length ($L=\frac{2\beta}{\kappa} r$) and cross-sectional radius (r) of the FR will increase during its expansion and propagation throughout the interplanetary space (see the expression of $B_0$ in Equation \ref{axial_field_strength}). Assuming that the angular width ($2\beta$) of the CME remains constant throughout its propagation and the nature of expansion is self-similar, we can consider that $L \propto r$. Therefore, considering the conservation of magnetic flux ($\phi_p= constant$), the axial magnetic field-strength (B$_0$) of any FR having a cross-sectional radius r will follow the relation

\begin{equation}
  B_{0} \propto \frac{1}{r^2}
  \label{b1}
\end{equation}

Thereby, knowing the cross-sectional radius $R_t$ ($R_i\le R_t\le R_f$) of the FR during its passage through the spacecraft using the Equation \ref{x4}, we can estimate its axial field-strength ($B_{t}$) at any time t ($0\le t\le t_p$) using the following relation

\begin{equation}
  B_{t}=B_{0_{CME}}\times{\frac{{r_{CME}}^2}{{R_t}^2}}
  \label{b2}
\end{equation}

where, $r_{CME}$ is the cross-sectional radius and  $B_{0_{CME}}$ is the axial magnetic-field strength of the near-Sun FR obtained from the observations as discussed in Section \ref{first}.

As the spacecraft intersects the MC along the path AB (see Figure \ref{expansion}), at any location ($x$) along AB the magnetic field vectors of the FR can be obtained using a cylindrical flux rope solution (Lundquist 1950) in a local cylindrical coordinate ($r, \phi, z$) attached to the MC axis. The magnetic vectors in the aforementioned ($r, \phi, z$) coordinate system will be

\begin{equation}
  B_{r}=0
  \label{f1}
\end{equation}

\begin{equation}
  B_{\phi}=H\times{B_{t}\times{J_1(\alpha{r})}}
  \label{f2}
\end{equation}

\begin{equation}
  B_{z}={B_{t}\times{J_0(\alpha{r})}}
  \label{f3}
\end{equation}

where, H=$\pm$1 is the handedness or sign of the helicity which is same as that of the near-Sun FRs according to the conservation of helicity rule,  $\alpha$ is the constant force-free factor, and $J_0$ and $J_1$ are the Bessel functions of order 0 and 1, respectively. The boundary of the FR is located at the first zero of $J_0$, which leads to $\alpha=\frac{2.41}{R_t}$ and $R_t$ is therefore the radius of the flux rope. $B_t$ and $R_t$ evolve according to the relation described in Equations \ref{b2} and \ref{x4} respectively. 

As we have assumed that after 10 $R_S$ the CME does not suffer any significant rotation and deflection, therefore the final elevation angle ($\theta$) of the MC axis at 1 AU should follow the tilt angle ($\delta$) of the CME and the azimuthal angle of the MC should follow the propagation longitude ($\phi$) of the CME obtained from the GCS fitting as discussed in Section \ref{sec:gcs}. In order to get the final magnetic field vectors in Geocentric Solar Ecliptic (GSE) coordinate system \citep{Hapgood}, we first transform the $B_r$, $B_{\theta}$ and $B_{\phi}$ from the local cylindrical coordinate ($r, \theta, z$) to the local cartesian coordinate ($x^{'}, y^{'}, z^{'}$) attached to the MC axis. Thereafter, knowing the azimuthal ($\phi$) and elevation ($\theta$) angle of the MC axis we transform the magnetic vectors $B_{x^{'}}, B_{y^{'}}$ and $B_{z^{'}}$ from the local cartesian coordinate ($x^{'}, y^{'}, z^{'}$) to the GSE coordinate system (x, y, z). Thus, we get the predicted magnetic vectors $B_x$, $B_y$ and $B_z$ of the ICME as detected by the spacecraft at 1 AU. 

\begin{figure*}[!t]
\centering
\includegraphics[width=.83\textwidth]{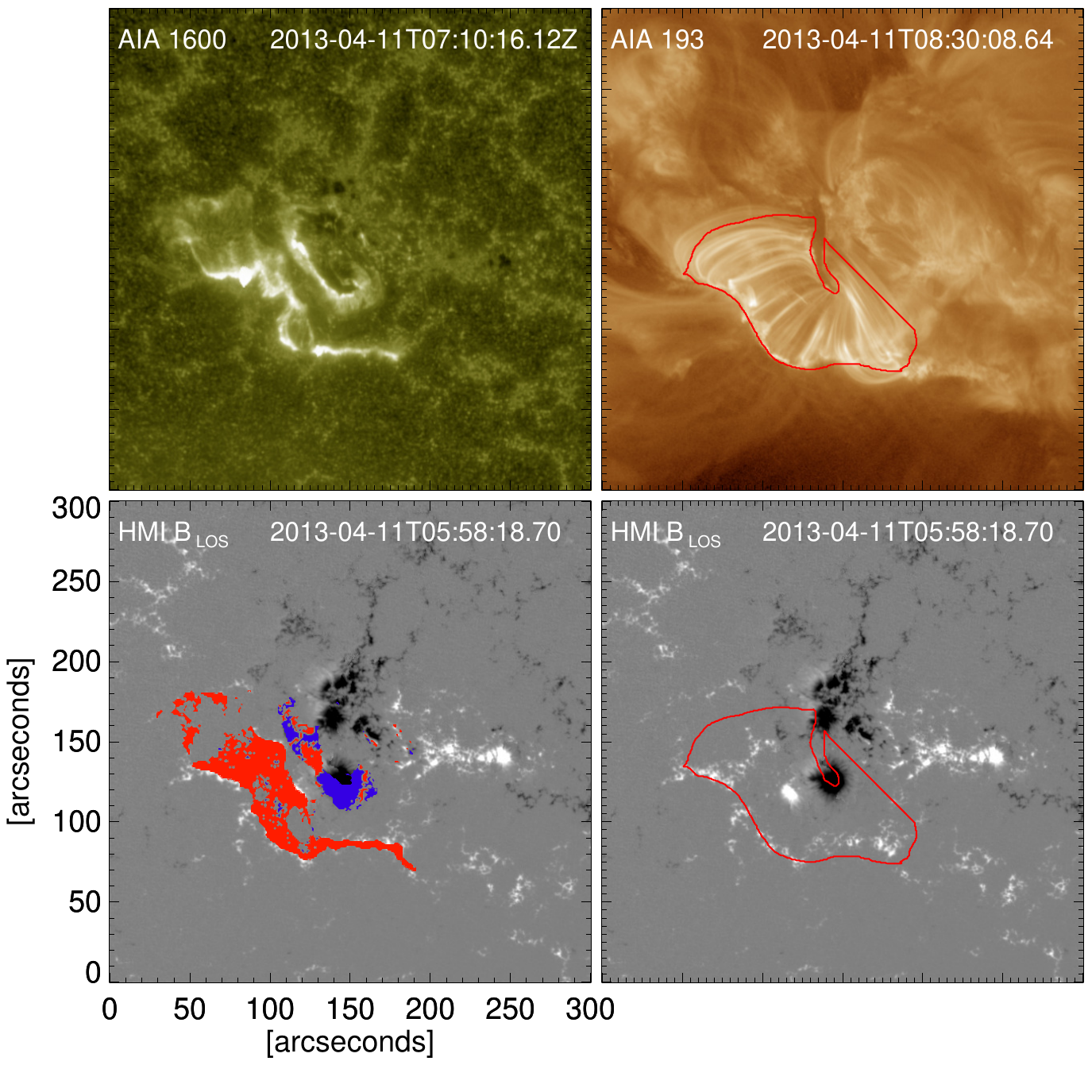}
\caption{Upper-left panel depicts the flare ribbon in AIA 1600 \AA\ image. The red boundary line in upper-right panel marks the post eruption arcades (PEAs) in AIA 193 \AA\ image. Lower-left and lower-right panel illustrate the HMI line-of-sight magnetic field. The red and blue regions in lower-right panel depict the cumulative flare ribbon area overlying the positive and negative magnetic field respectively. The red boundary in lower-right panel
is the over-plotted PEA region.}
\label{recconection_flux}
\end{figure*} 

\section{INFROS Model validation: A test case for the CME event on 2013 April 11}\label{model_validation}

 \begin{figure*}[!t]
\centering
\includegraphics[width=1\textwidth]{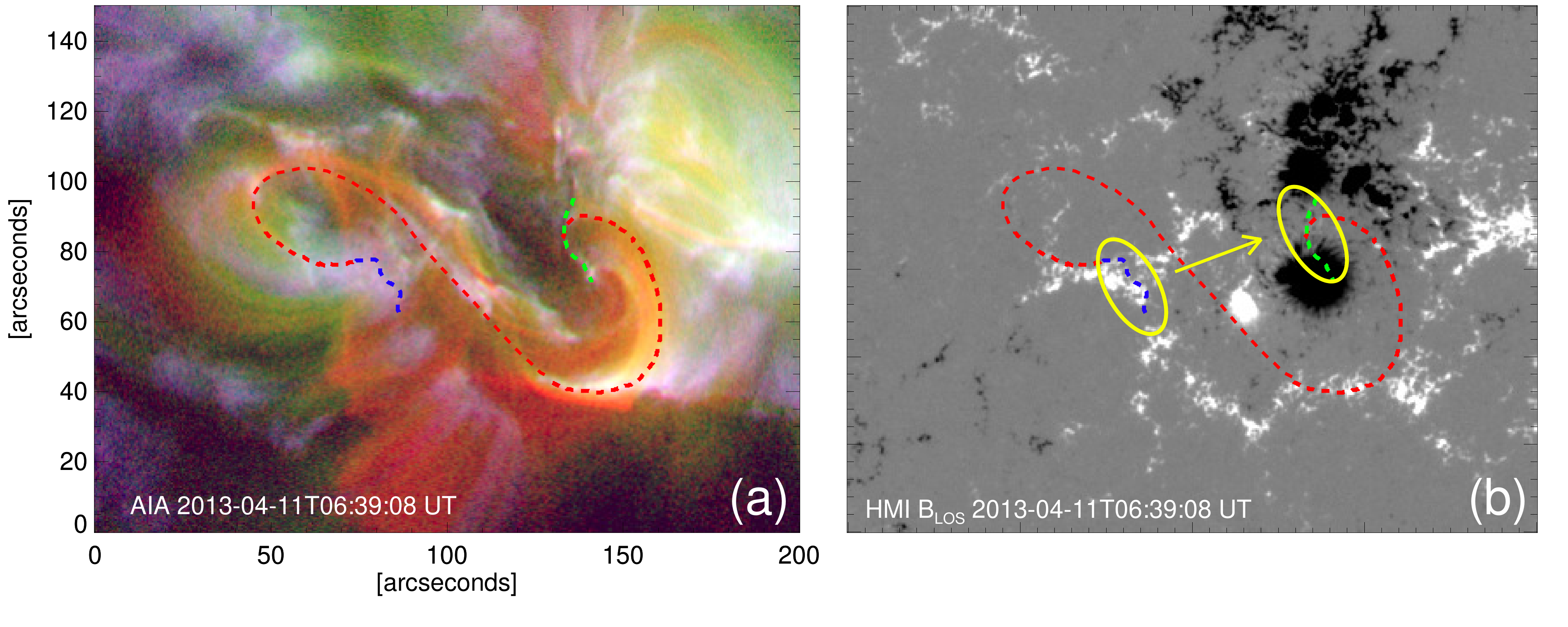}
\includegraphics[width=1\textwidth]{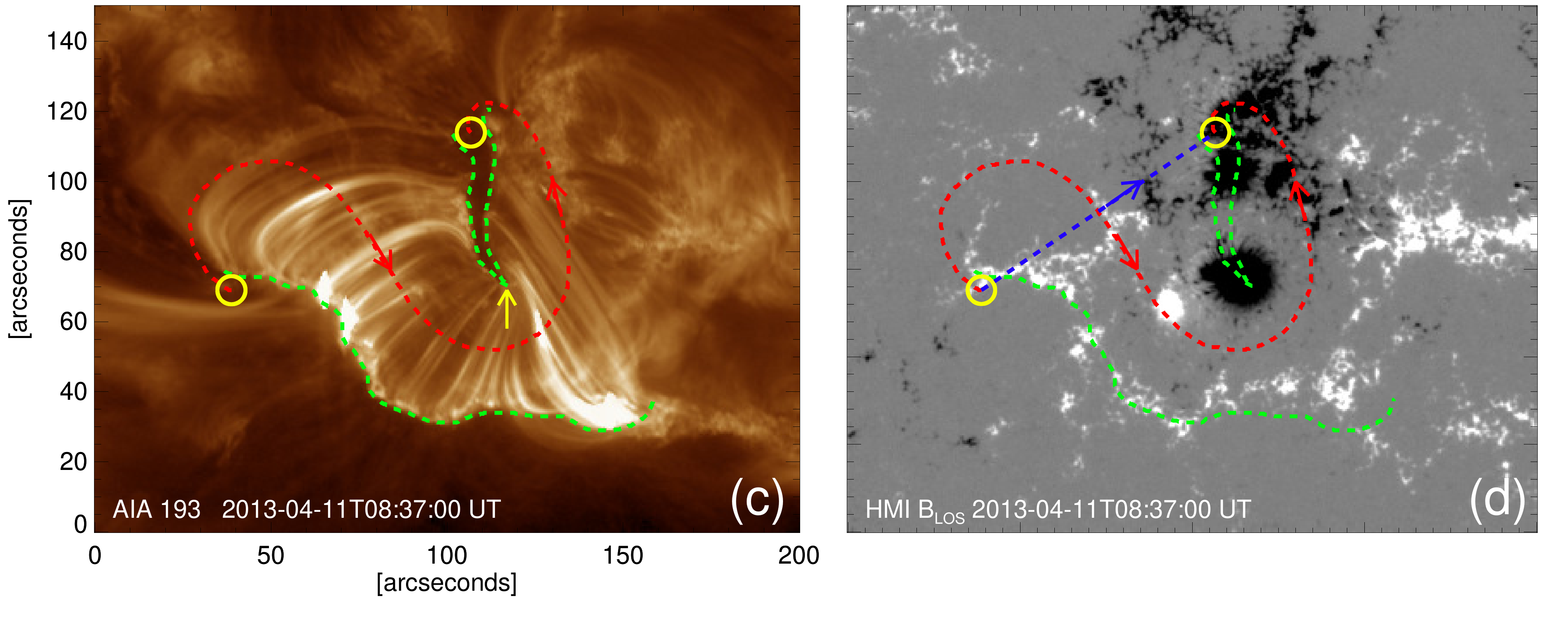}
\caption{The pre-flare sigmoidal structure observed in the composite images constructed from the AIA 94 \AA\ (red), 335 \AA\ (green) and 193 \AA\ (blue) passband observations (a). The associated HMI line-of-sight magnetic field plotted in gray scale within saturation values $\pm$ 500 G (b). The red dashed line (plotted in panel (a)) that approximately resembles the sigmoidal structure has been overlaid on the HMI line-of-sight magnetic field in panel (b). The blue and green dashed lines in panel (a) approximately denote the boundaries where the two ends of the bundle of sigmoidal field lines are rooted. The same blue and green dashed lines are overlaid in panel (b). The post-eruption arcades (PEAs) observed in AIA 193 \AA\ passbands (c) and the associated HMI line-of-sight magnetic field (d). The green-dashed lines in panel (c) mark the two side boundaries of the PEA and the same is overlaid in panel (d). The red dashed line is drawn along the approximate center of the two side boundaries of the PEA, connecting the two expected foot-point locations (shown by the yellow circles) of the erupting flux rope. The blue-dashed line connecting the flux rope foot-points and the blue-arrow in panel (d) indicate the north-west direction.}

\label{sigmoid}
\end{figure*}

\begin{figure*}[!t]
\centering
\includegraphics[width=1\textwidth]{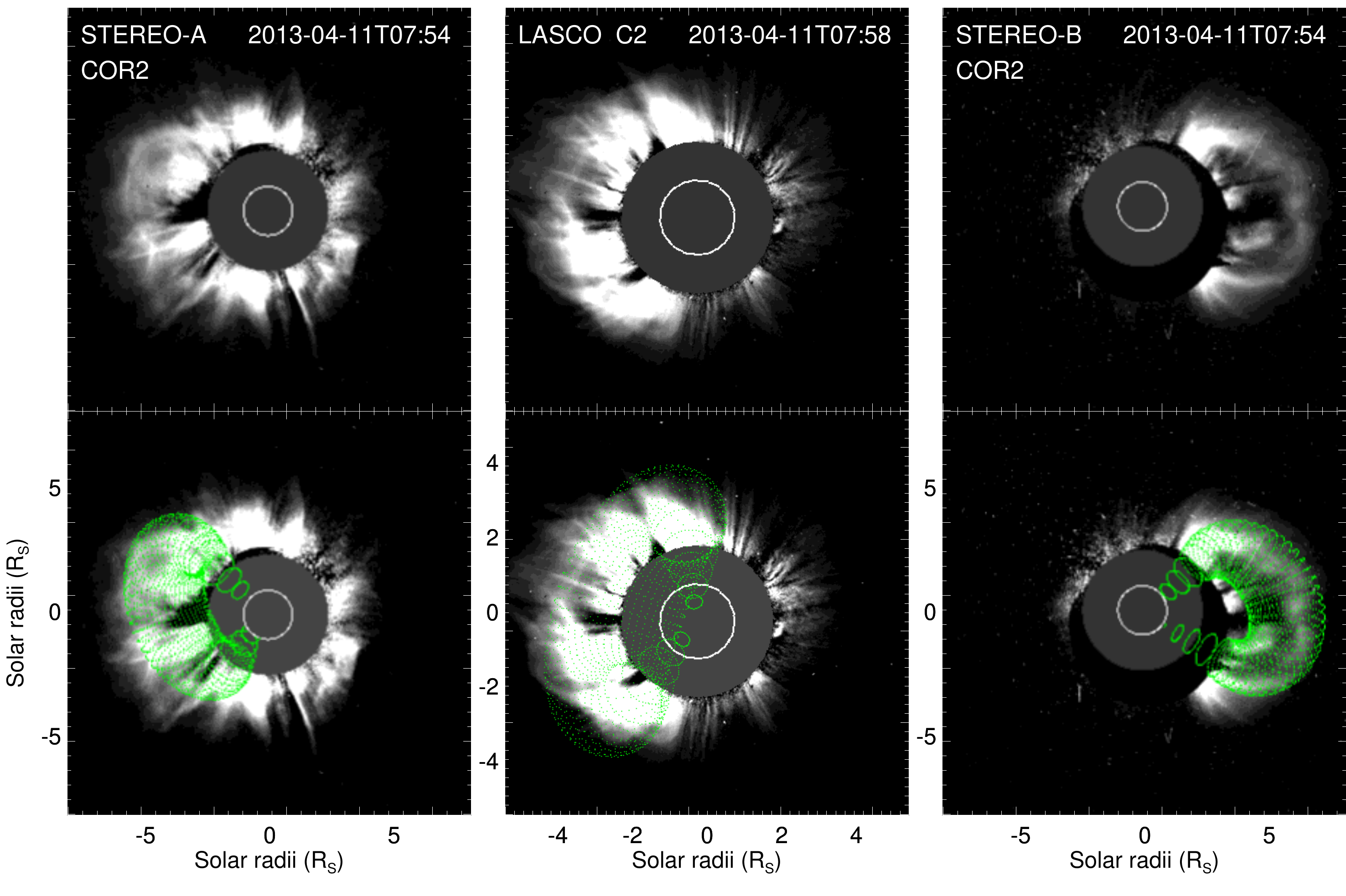}
\caption{Top panels depict the CME morphology observed in COR2-A (top-left), LASCO C2 (top-middle), and COR2-B (top-right), respectively,
at 07:54 UT on 2013 April 11. Bottom panels illustrate the overplot of the best-fitted wire frame (green dotted marks) of the FR using the GCS model.}
\label{gcs_fit}
\end{figure*}

As a proof of concept we validate our model (INFROS) for an Earth-directed CME which erupted from the Sun on 2013 April 11 at around 06:50 UT. The CME was associated with an M6.6 class solar flare  \citep{Cohen_2014,Lario_2014,Vema1,Vema2,Joshi_2016,Fulara2019} that occurred in the active region (AR) 11719. Its arrival at the L1 point was detected with the signature of shock arrival on 2013 April 13 at 22:54 UT, FR leading edge on 2013 April 14 at 17:00 UT and a trailing edge on 2013 April 15 at 19:30 UT. The smooth variation and rotation in its in-situ magnetic field profile along with the low proton temperature hold the characteristic signatures of an MC \citep{Burlaga}. Moreover, the CME did not exhibit any interaction with other CMEs and evolved as an isolated magnetic structure from the Sun to Earth. Therefore, the basic assumptions made in our model hold good for this case study.

The evolution of the flare ribbons and the formation of post eruption arcades (PEAs) associated with the M6.6 class flare (see figure \ref{recconection_flux}) were well observed by the  Atmospheric Imaging Assembly (AIA) \citep{lemen} and the Helioseismic and Magnetic Imager (HMI) \citep{Schou} onboard Solar Dynamics Observatory (SDO) \citep{Pesnell}. Furthermore, the multi-vantage point observations from STEREO-A, STEREO-B and LASCO were suitable to reconstruct the 3D morphology of the associated CME. Therefore, we are able to determine all the near-Sun FR properties of the CME in order to use those as realistic  inputs for INFROS model.

\subsection{Model inputs for the CME event on 2013 April 11}\label{inputs}

\subsubsection{Poloidal flux content of the flux-rope}\label{p_flux} We calculate the flare associated reconnection flux by applying both the methods \citep{Longcope,Qiu,Gopalswamy1} as described in Section \ref{fred}. The red and blue regions in the lower left panel of figure \ref{recconection_flux} show the cumulative flare ribbon area overlying the positive and negative polarities of photospheric magnetic field respectively. The average of the absolute values of positive and negative magnetic fluxes underlying the cumulative flare ribbon area yield the value of reconnection flux as $1.9 \times 10^{21}$ Mx. Taking into account the formation-height of the flare ribbons, we have incorporated a 20$\%$ correction \citep{Qiu} in the estimation of reconnection flux. The half of the total unsigned magnetic flux underlying the PEA (the region enclosed by the red boundary as shown in upper-left and lower-left panels of figure \ref{recconection_flux}) yield the value of reconnection flux as $2.3 \times 10^{21}$ Mx.  In order to determine the magnetic properties of the associated CME we equate the poloidal flux content of the FR to the average value ($2.1 \times 10^{21}$ Mx) of the reconnection fluxes obtained from the aforementioned two methods.

\subsubsection{Direction of the axial-magnetic field and the chirality of the flux-rope} 
The source location of the M6.6 flare that occurred in AR 11719, was associated with a pre-eruptive sigmoidal structure \citep{Vema2,Joshi_2016}. Panel (a) of Figure \ref{sigmoid} shows the highly skewed pre-flare sigmoid observed in EUV images of AIA passbands (94 \AA, 335 \AA\ and 193 \AA). The observed inverse S-shaped morphology of the sigmoidal structure (indicated by the red dashed line) has been overlaid on the HMI line-of-sight magnetogram (panel (b) of Figure \ref{sigmoid}), which reveals the left handed chirality of the associated flux-rope. This follows the hemispheric helicity rule \citep{Bothmer} as the source region of the CME was located in the northern solar hemisphere.

We identify the two boundaries as shown by the blue and green dashed lines in panel (a) of Figure \ref{sigmoid}, where the two ends of the bundle of sigmoidal field lines are rooted during the pre-eruptive phase. The two aforementioned boundaries are overlaid on the HMI line-of-sight magnetic field and the regions are marked by the yellow ellipses (see panel (b)). The simple connectivity (without considering any twist) between the two opposite magnetic polarities underlying the regions marked by the yellow ellipses suggests the north-west direction (as shown by the yellow arrow) as the axial orientation of the FR at higher heights in the corona (above $\approx$ 5 $R_S$). This is expected as the apex-orientation of the left-handed FR should rotate in counter-clockwise direction to release the axial twist or writhe during its evolution in the lower corona below 5 $R_S$ \citep{Lynch}.
 
In order to confirm the axial orientation of the FR, we further investigate the morphology of the associated PEA formed during the flare. Panel (c) of Figure \ref{sigmoid} shows that the eastern part of the PEA channel is tilted towards the south-west direction and further bends towards the north-west direction at the location indicated by the yellow arrow, forming a nearly U-shaped morphology. This is certainly a complex morphology which makes the event  more complicated. Considering the apex orientation of the FR inferred only from the eastern part of the PEA channel, \citet{Palmerio2018} found contradiction between the solar and 1-AU Bz direction. However, we have focused on the full U-shaped morphology of the PEA channel in this study. Considering the full extent of the PEAs allows us to analyze the FR structure beyond the sigmoidal pre-eruptive configuration and, therefore, to capture the complete evolution of the FR in the lower corona during the phase of sigmoid to arcade formation. According to the standard flare model in three dimension \citep{Shibata_1995,Moore,Pf2002}, the foot-points of the eruptive FRs are believed to be located on either side of the two ends of the PEA channel. Therefore, considering the left-handed chirality, we mark the expected locations of the two foot-points of the FR as shown by the yellow circles at the two ends of the U-shaped PEA channel. The red dashed curve connecting the two yellow circles indicates the possible writhe presented in the FR during the formation phase. This is in agreement with the observed writhing motion of that FR during the eruptive phase as reported by \citet{Joshi_2016}. Therefore, due to the writhing motion the FR would have relaxed the axial-twist during its evolution in the lower corona, resulting in an orientation following the straight connectivity (shown by the blue dashed line in panel (d) of Figure \ref{sigmoid}) between the two foot-point locations. In such scenario, the magnetic polarities underlying the two yellow circles clearly indicates that the axial orientation of the FR is directed towards north-west.

From the GCS fitting (Figure \ref{gcs_fit}) of the observed white-light morphology of the CME at $\approx$ 10 $R_S$, we estimate the tilt angle of the CME axis as $73 \pm 10^{\circ}$ with respect to the ecliptic plane. Minimizing the difference in angle between the GCS tilt and the axial direction (north-west) of the FR inferred from the on-disk observations, we obtain the axial magnetic-field direction of the CME FR at $\approx$ 10 $R_S$ along $73\pm 10^{\circ}$, measured in counter-clockwise direction with respect to the solar equator. Assuming that no major rotation occurred after 10 $R_S$, we consider this axis orientation as the final orientation of the associated MC axis at 1 AU.

\begin{figure*}[!t]
\centering
\includegraphics[width=1\textwidth]{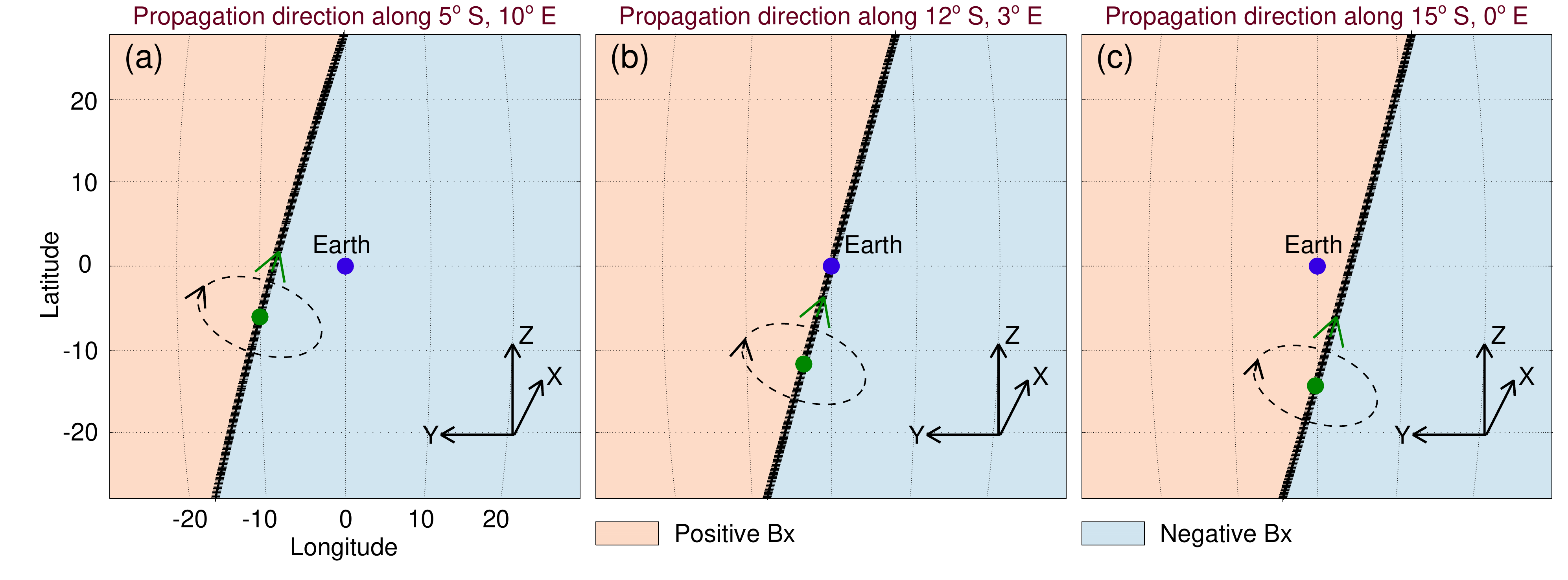}
\caption{Location of Earth (blue dots) with respect to the magnetic axis (black solid lines) of the CME projected on the solar disk. The green dots in each panels show the three different propagation direction of the CME. The black and green arrows denote the direction of poloidal and axial magnetic field of the flux-rope respectively. Any virtual space-craft that resides at the left/right side of the magnetic axis (denoted by the pink and blue shaded region respectively) will encounter the $B_x$ component of the flux-rope as positive/negative in GSE coordinate system.}
\label{b_x}
\end{figure*}

\subsubsection{Axial field-strength of the flux-rope}
In order to estimate the axial field-strength of the near-Sun FR we first determine the geometrical parameters associated with it. The top panels of figure \ref{gcs_fit} show the white-light morphology of the CME as observed in base difference images obtained from STEREO-A/B and LASCO. The GCS fitting (bottom panels of figure \ref{gcs_fit}) to the multi-vantage point observations of the CME yields the aspect ratio ($\kappa$) and the half-angular width ($\beta$) of the CME as 0.22 and 26$^{\circ}$ respectively. Therefore, the length ($L=2\beta R$) of the associated FR at a radial distance (R) of 10 $R_S$ is estimated as approximately 9 $R_S$. Using Equation \ref{axial_field_strength}, we obtain the axial field-strength of the FR at 10 $R_S$ as 52 mG.

\subsubsection{Propagation direction of the CME}\label{direction}
The GCS fitting (Figure \ref{gcs_fit}) of CME morphology at $\approx$ 10 $R_S$ yields the propagation direction of the CME along S05E10. Taking into account an uncertainty of 10$^{\circ}$ in determining both the longitude and latitude of propagation direction, we have performed the GCS fitting several times and found the propagation direction of the CME to lie within the range $0 -10^{\circ}$ E and $5-15^{\circ}$ S. Using the range of values of the propagation direction and the tilt angle ($73\pm 10^{\circ}$) of the CME as inputs, we estimate the impact distance of the CME magnetic axis at 1 AU within the range 0 to 21 $R_S$.\\

  \begin{figure*}[!t]
\centering
\includegraphics[width=\textwidth]{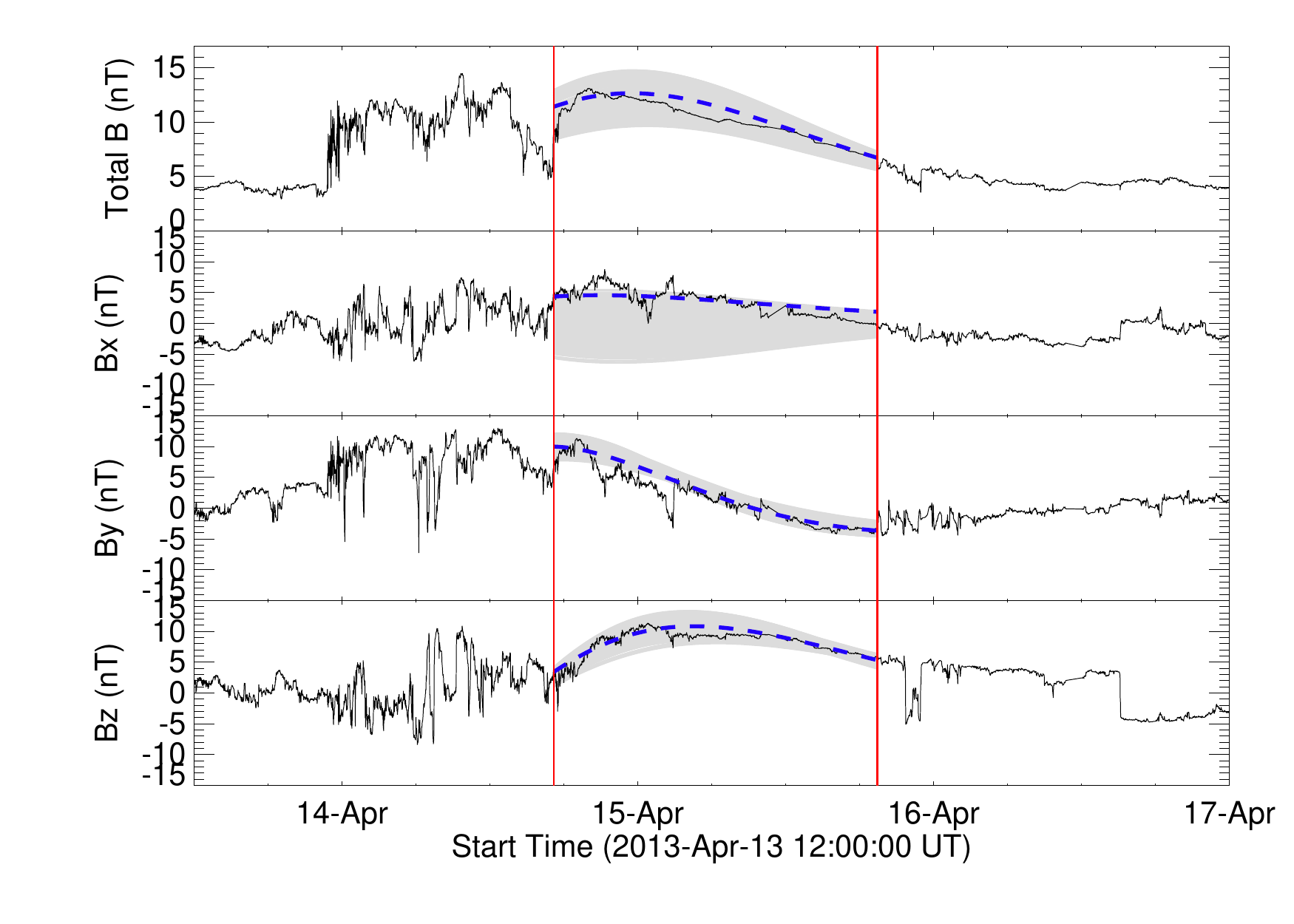}
\caption{Magnetic vectors as detected by the WIND spacecraft for 2013 April 14 ICME event. The two red vertical lines denote the magnetic cloud boundary. The blue dashed lines denote the predicted magnetic vectors obtained from the model which best match the observed magnetic profiles of the MC. The gray shaded regions denote the uncertainty in predicting the respective magnetic vectors.}
\label{model_plot}
\end{figure*}

\subsection{Sensitivity of the estimated magnetic vectors to the propagation direction and tilt angle of the CME}\label{sensitivity}

We notice that the sign of $B_x$ component for the estimated magnetic vectors of the ICME as detected by any spacecraft aligned along Sun-Earth line is very sensitive to the propagation direction of the CME. The three panels in figure \ref{b_x} depict the location of Earth (denoted by blue dots) with respect to the magnetic axis (denoted by black solid lines) of the CME propagating along three different directions which are within the error limits as estimated in Section \ref{direction}. In each of the three panels the Sun-grids are shown within $\pm$ 30$^{\circ}$ longitude and latitude where the projected location of Earth on the solar disk resides at 0$^{\circ}$ longitude and 0$^{\circ}$ latitude. Keeping the tilt angle as 73$^{\circ}$ we project the magnetic axis of the CME on the solar disk as shown by the black solid lines in each panel. The green dots and arrows on the magnetic axis denote the propagation direction of the CME and the direction of axial magnetic field of the associated FR respectively. The arrows along the black dashed lines surrounding the CME magnetic-axis depict the direction of poloidal magnetic field according to the left-handed chirality of the associated FR.

Notably, at any projected location on the solar disk which lies on the left/right side of the CME axis, the direction of poloidal magnetic field will be towards/outwards the Sun. Accordingly, the sign of $B_x$ will change at any location on the either side of the magnetic axis which we have shown by the pink and blue regions where $B_x$ possesses positive and negative values respectively. Panel (a) in figure \ref{b_x} shows that the projected location of the Earth lies on the region of negative $B_x$ for the estimated direction (10$^{\circ}$ E, 5$^{\circ}$ S) and tilt (73$^{\circ}$ with respect to the ecliptic plane) of the MC axis as obtained from GCS fitting. However, a small shift in the propagation direction from 10$^{\circ}$ E, 5$^{\circ}$ S to 3$^{\circ}$ E, 12$^{\circ}$ S results in a zero impact distance between the MC axis and the Sun-Earth line (see panel (b) in figure \ref{b_x}) for which the estimated $B_x$ component turns out to be zero. If we further shift the propagation direction of the MC axis from 3$^{\circ}$ E, 12$^{\circ}$ S to  0$^{\circ}$ E, 15$^{\circ}$ S within the error limits, the sign of $B_x$ becomes positive as the location of Earth or any spacecraft aligned along Sun-Earth line lies on the left side of the MC axis where the direction of poloidal magnetic field is towards the Sun (see panel (b) in figure \ref{b_x}).   Therefore, our analysis shows that within the error limits of the propagation direction of the CME, $B_x$ can have both positive and negative components in the estimated magnetic vectors of the ICME at 1 AU.
 
\begin{figure*}[!t]
\centering
\includegraphics[width=\textwidth]{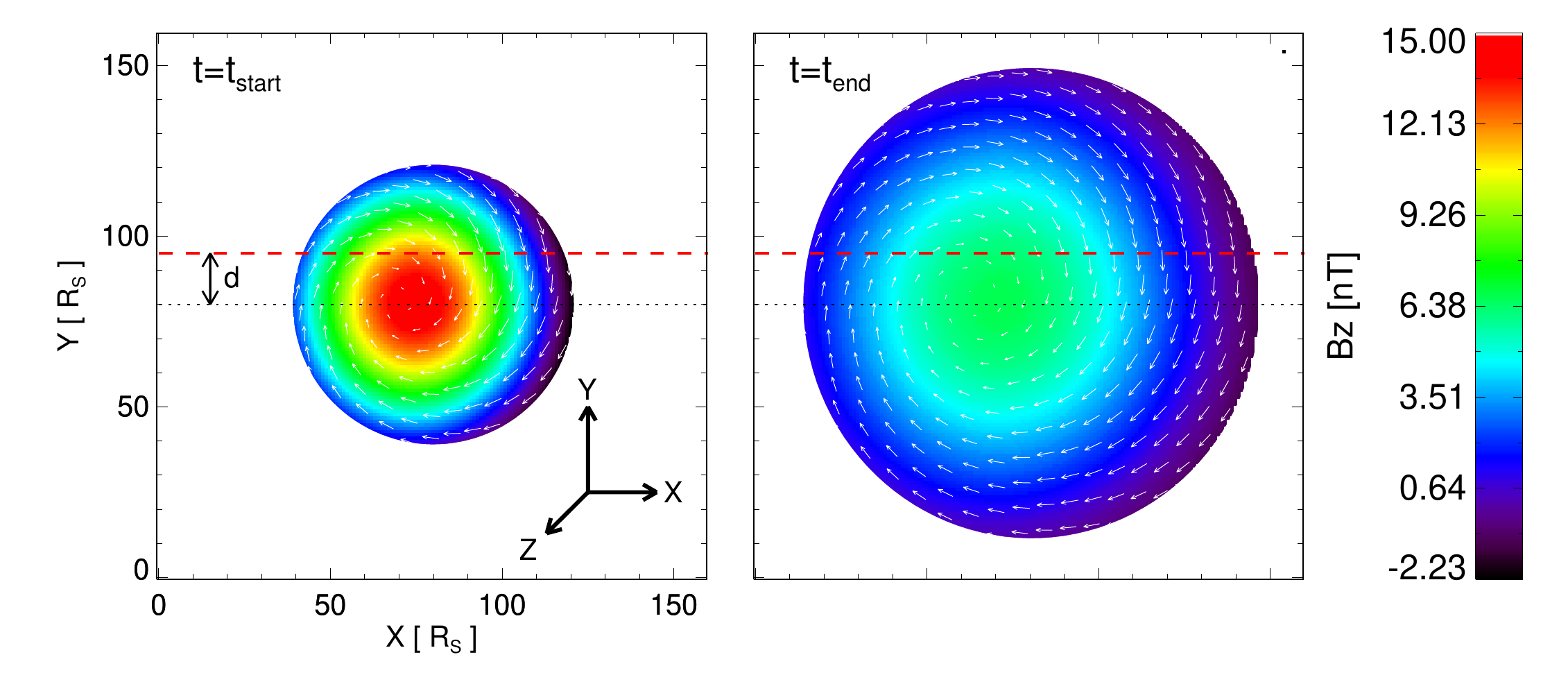}
\caption{Left panel: Cross-section of the flux rope as viewed on the ecliptic plane when the spacecraft at 1 AU just encounters the arrival of the magnetic cloud (MC). Right panel: Cross-section of the expanded flux-rope when the spacecraft completes its passage through the magnetic cloud and reaches to the rear boundary of the MC at 1 AU. The color bar shows the strength of the southward component of magnetic magnetic field in GSE coordinate which is positive outwards the plane of the paper. The black dotted line passes through the axis of the flux-rope. The white arrows mark the direction of magnetic field component lies on the ecliptic plane inside the MC boundary. The red dashed line at a distance d (impact distance) from the black dotted line shows the spacecraft trajectory along which the MC is intersected by it at 1 AU.}
\label{impact_trajectory}
\end{figure*}

  It is noteworthy that the above mentioned scenario is true for any tilt angle of the FR orientation where the propagation direction is very close to the Sun-Earth line. Interestingly, the sign or the direction of variation (positive to negative or vice-versa) of the estimated $B_y$ and $B_z$ components are not sensitive to the small variations ($\pm$10$^{\circ}$) in the propagation direction and tilt angle of the CME. Therefore, we expect less uncertainty in the prediction of $B_y$ and $B_z$ components of the MC.    
\subsection{Model outputs}

Using the near-Sun FR properties of the associated CME as described in Section \ref{inputs}, we estimate the magnetic vectors of the ICME as intersected by the spacecraft at 1 AU. The curves shown by the black solid lines in figure \ref{model_plot} depict the observed magnetic vectors of the ICME as detected by the WIND spacecraft \citep{OGILVIE}. The red vertical lines denote the front and rear boundary of the MC which we have estimated from the observed magnetic field and plasma parameters of the ICME.

Incorporating the uncertainties in the GCS parameters involved in the modeling, we generate all the possible input data-sets from the range of values of the input parameters, i.e. the propagation direction ($0 -10^{\circ}$E, $5-15^{\circ}$ S), tilt-angle ($63-83^{\circ}$) and aspect-ratio ($0.20-0.24$) of the CME. Further, considering an error of $2\times10^{20}$ Mx (standard deviation of the two values of reconnection flux obtained from the two different methods as discussed in Section \ref{p_flux}) in determining the poloidal flux and $\pm\ 0.02$ in determining the CME aspect-ratio, we get 20$\%$ error in estimating the axial field strength ($B_{0_{CME}}$) of the CME. This yields the estimated range of $B_{0_{CME}}$ at 10 $R_s$ as $42-62$ mG, with a mean value of 52 mG. This is consistent with the average value of the distribution of axial fields at 10 $R_S$ \citep{Gopalswamy2018}. Using these sets of input data we run our model and generate synthetic magnetic profiles of the MC. Among these sets of predicted magnetic vectors, we find that the magnetic profiles (shown by the blue dashed lines in each panels of figure \ref{model_plot}), which best match the observed magnetic vectors of the MC, can be obtained by using the propagation direction along $0^{\circ}$E, $15^{\circ}$ S, the tilt angle as $73^{\circ}$, the aspect-ratio as 0.22 and the axial field strength at 10 $R_s$ as 52 mG. For this set of input parameters, we show the spacecraft trajectory through the MC and the magnetic field profiles of the FR cross-section when the spacecraft intersects the front and rear boundary of the MC
(Figure \ref{impact_trajectory}). The uncertainty in predicting the magnetic vectors as shown by the gray shaded region in each panel of figure \ref{model_plot} is obtained by overplotting all the sets of output magnetic profiles.

\begin{figure*}[!t]
\centering
\includegraphics[width=1\textwidth]{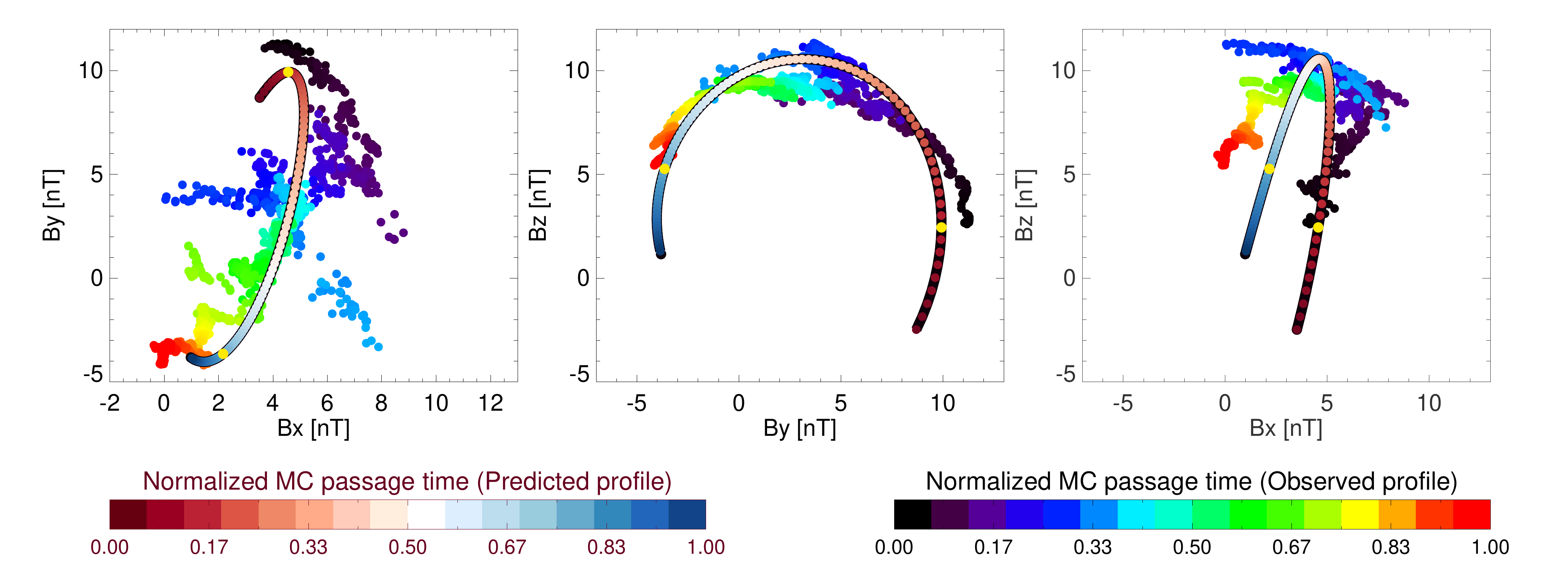}
\caption{Hodogram plot of the magnetic field
vectors within the MC for both observed and modeled data values. The yellow dots drawn over the plots for modeled data values denote the data points which approximately match the front and rear boundary of the observed MC.}
\label{hodogram}
\end{figure*}

In order to overplot the modeled magnetic vectors within the temporal window of the observed MC, we identify the front and rear boundary of the modeled MC from the hodogram analysis. Figure \ref{hodogram} shows the scattered plots among the magnetic field vectors within the MC for both observed and modeled data values. The yellow dots drawn over the plots for modeled data values denote the data points which approximately match the front and rear boundary of the observed MC. Therefore, we take the observed MC boundary as a reference boundary and overplot the data-points of the modeled magnetic vectors which lie in between the two yellow dots.

Figure \ref{model_plot} shows that the predicted  magnetic field profiles of the MC obtained from our model are in good agreement with those of the observed profiles as detected by the WIND spacecraft. In comparison to the $B_y$ and $B_z$ components, the larger uncertainty arisen in predicting the $B_x$ component is due to its sensitivity towards the propagation direction of the CME which we have discussed in Section \ref{sensitivity}. Nevertheless, the predicted profiles for $B_y$ and $B_z$ components show good  agreement with the observed profiles. The predicted strength of of the $B_z$ component has been found to be $10.5 \pm 2.5$ nT when the MC axis makes its closest approach to the spacecraft. This is in agreement with the  maximum observed strength (11 nT) of the $B_z$ component obtained from the in-situ data. Therefore, our model successfully predicts both the strength and the general profile of the $B_z$ component of the MC with a good accuracy. \\ 

\section{Conclusion}\label{discussion}
We have presented an analytical model (INFROS) to predict the magnetic field vectors of ICMEs based on realistic inputs obtained from near-Sun observations. As a proof of concept, we validate our model for the 2013 April 11 CME event. The predicted magnetic field-vectors of the ICME obtained from INFROS show good agreement with those observed by the WIND spacecraft at 1 AU. This shows promising results in forecasting of Bz in real time.

There are several key aspects in which INFROS appears to be superior than the existing semi-analytical \citep{2017Kay} and analytical \citep{Savani2015} models. The analytical model proposed by \citet{Savani2015} does not incorporate the expanding nature of the ICME during its passage through the spacecraft which yields an unrealistic symmetric profile of the total magnetic field strength of the ICME with time. \citet{2017Kay} included the expanding nature of ICMEs in their semi-analytical model using the speed and duration of passage of the ICME measured at 1 AU as free parameters. However, the formulation developed in INFROS incorporates the FR expansion in such a way so as to get rid of the unknown parameters like the expansion speed (V$_{exp}$), propagation speed (V$_{pro}$) and the time of passage (t$_{p}$) of the ICMEs at 1 AU (see section \ref{SSE}). Moreover, none of the existing models \citep{Savani2015, Kay2017,Mostl} were capable of predicting the time-varying axial field strength of the expanding flux-rope embedded in ICMEs during its passage through the spacecraft. Therefore, it was not possible to forecast the strength of the southward component of magnetic field (Bz) embedded in the ICMEs in order to predict the severity of the associated geomagnetic storms. It is worth noting that INFROS is capable of predicting the time-varying axial field strength and the expanding nature of the interplanetary FR without involving any free parameters, as all the input parameters are constrained either by the near-Sun observations or the inherent assumptions (self-similar expansion) made in the model. Therefore, the modeling approach proposed in this article turns out to be a promising space-weather forecasting tool where the magnetic field vectors of the ICMEs can be predicted well in advance using the near-Sun observations of CMEs.

In order to reduce the uncertainties involved in the model predictions, INFROS can be further constrained by the inputs obtained from the spacecraft orbiting at different heliocentric distances in between Sun and Earth (e.g. MESSENGER, VEX, Parker Solar Probe etc.). In a future study, we plan to validate this model for the ICMEs detected by multiple spacecraft orbiting at different heliocentric distances, which will give better insight into the magnetic field variation from the Sun in the direction of the spacecraft.

We thank the referee for helpful comments
that improved the quality of this manuscript. This work was performed at the NASA Goddard Space Flight Center under the aegis of the SCOSTEP Visiting Scholar program (RS). RS thanks SCOSTEP and NASA (via the Catholic University of America) for the financial support. NG was partly supported by NASA's Living with a Star program.

\bibliographystyle{yahapj}

\end{document}